\documentclass{article}
\usepackage{ijcai20}

\usepackage{times}

\usepackage{soul}
\usepackage{url}
\usepackage[hidelinks]{hyperref}
\usepackage[utf8]{inputenc}
\usepackage[small]{caption}
\usepackage{graphicx}
\usepackage{amsmath}
\usepackage{booktabs}
\title{Pilot: Winner of the Human-Agent Negotiation Challenge at IJCAI 2020}

\author{
Kushal Chawla, Gale Lucas\\
\affiliations
University of Southern California, Los Angeles, USA\\
\emails
\{chawla,lucas\}@ict.usc.edu
}

\begin{document}
\maketitle
\begin{abstract}
This document describes our agent \textbf{Pilot}, winner of the Human-Agent Negotiation Challenge at ANAC, IJCAI $2020$. \textbf{Pilot} is a virtual human that participates in a sequence of three negotiations with a human partner. Our system is based on the Interactive Arbitration Guide Online (IAGO) negotiation framework. We leverage prior Affective Computing and Psychology research in negotiations to guide various key principles that define the behavior and personality of our agent.
\end{abstract}
\section{Introduction}
Negotiation is integral to our everyday interactions, be it in a vegetable market, at legal proceedings, or even while finalizing business deals. Negotiation is a complex skill, requiring the ability to combine reasoning with linguistic and affective knowledge, making it a challenging task for an automated system. While most of the research efforts in building automated negotiating agents have focused on agent-agent negotiations~\cite{Genius}, there is recent interest in building agents that can negotiate with humans as well~\cite{gratch2015negotiation}. Use-cases include pedagogy~\cite{johnson2019intelligent} and mediation~\cite{chalamish2012automed}. Development in this area will also be useful in advancing the capabilities of existing conversational assistants like Google Assistant, Alexa, and Siri. The Human-Agent Negotiation challenge at the Automated Negotiating Agents Competition (ANAC 2020) is one such step. In this document, we describe our submission for this competition.

Our agent, referred to as \textbf{Pilot}, is based on the Interactive Arbitration Guide Online (IAGO) platform~\cite{mell2016iago}. Through a combination of behavior, message, and expression policies, our agent negotiates with a human partner in a sequence of three back-to-back negotiations (described in Section \ref{sec:design}). We build on the baseline agent provided to us by the competition organizers\footnote{\url{https://myiago.com/}}.

\section{Agent Design}
\label{sec:design}
In this section, we present key design features of \textbf{Pilot} and wherever possible, compare them with the baseline agent. We first describe the overall personality of \textbf{Pilot}, followed by key elements of the three agent policies, namely, behavior, message, and expression.

\subsection{Agent Personality}
Justifying the name, \textbf{Pilot} attempts to lead the negotiation with the human partner, while also maintaining a friendly persona. This allows \textbf{Pilot} to cater to both the objectives of high performance and positive opponent perception.

To achieve this, our agent pushes the human partner to first share some of their preferences before discussing any offers. Previous results at ANAC show that humans tend to remain truthful and send out only a few offers themselves~\cite{mell2018results}. Hence, this push allows the agent to build a reasonable opponent model before rolling out offers. Further, the agent uses phrases like "Let me help you out" to help the human in navigating the negotiation and as a result, portrays itself as experienced and builds a rapport with the human partner, which has been widely shown to build the joint value in negotiations~\cite{nadler2003rapport}. Further, the agent does not lie about its preferences and does not hold back information, when explicitly asked. However, it also does not give any explicit information, if not asked. 
\subsection{Behavior Policy}
\textbf{Pilot} follows a competitive negotiation strategy in all three negotiations, where it starts off with a high initial offer and keeps conceding least wanted items one by one. Since the agent pushes its partner to share their preferences first, it starts off with a more reasonable initial offer, which has a better chance of getting accepted than an all-vs-none offer. However, this initial offer does become more favorable for the agent in later negotiations.

\noindent\textbf{Only dealing in full offers}: Inspired from Agent Wotan~\cite{mell2018results}, \textbf{Pilot} only deals in full offers in order to save time.

\noindent\textbf{Indulging in favor exchange}: The baseline agent attempts to exchange favors early on in the negotiations. This has a couple of disadvantages: 
\begin{enumerate}
    \item First, whether the favor request is fruitful or not depends on the personality of the partner such as their Social Value Orientation~\cite{van1997development,cornelissen2009social}. Hence, this may not work with partners who portray competitive or selfish behavioral traits. In fact, based on prior ANAC competitions, favor exchange may even back-fire with selfish partners~\cite{mell2018towards}.
    \item Secondly, even if the favor request is accepted, the agent only takes minimal advantage by just claiming a single valuable item.
\end{enumerate}
Ideally, in order to fix the first problem, information about the personality traits of the partner can be useful. Unfortunately, such information is not available to the agent. Hence, \textbf{Pilot} leverages prior research which suggests that framing can help in promoting pro-social behavior~\cite{pulford2016social}. We use sentences such as "I am excited to build the value for both of us. I hope you are equally excited as well." and "Please remember: Our joint decisions will determine how many points we both earn." to prime the pro-social behavior in the human partners, with the aim of increasing the likelihood of the favor request being accepted.

In order to take the maximum advantage of favor acceptance, \textbf{Pilot} does not indulge in favor exchange early on, unlike the baseline agent. Instead, it holds off the request until the human partner has shared some of their preferences. If the favor is accepted, this allows our agent to roll out a full offer which greatly benefits the agent.

While returning the favor, the agent starts with a reasonable initial full offer based on the discussed preferences but further concedes three additional items to the human partner.
\subsection{Message Policy}
Our message policy mostly follows the provided baseline policy, with a few modifications:
\begin{enumerate}
    \item We modified several message choices in the baseline agent, which we believe might have confused the human partner in the context of the IAGO platform.
    \item If asked, the agent lies about its Best Alternative To a Negotiated Agreement (BATNA). \textbf{Pilot} inflates the original value of BATNA by $1.5$ times. This allows the agent to negotiate from a position of higher bargaining power.
\end{enumerate}
\subsection{Expression Policy}
The agent displays moderate expressions for the first two negotiations, promoting a healthy relationship with the partner. However, research has shown the benefits of extreme emotions such as anger in inducing more concessions~\cite{de2011effect}. Hence, for the third and final negotiation, the agent displays anger for offer rejects, in the hope of getting a better deal.
\section{Conclusion}
In this document, we describe the core elements of our virtual agent, \textbf{Pilot}. By strategically indulging in favor exchange after building a strong opponent model, \textbf{Pilot} builds the most value out of the negotiations, while still maintaining a positive perception in the eyes of the opponent, making it the winner of the negotiation challenge at IJCAI 2020.
\section{Acknowledgements}
Our research was sponsored by the Army Research Office and was accomplished under Cooperative Agreement Number W911NF-20-2-0053. The views and conclusions contained in this document are those of the authors and should not be interpreted as representing the official policies, either expressed or implied, of the Army Research Office or the U.S. Government. The U.S. Government is authorized to reproduce and distribute reprints for Government purposes notwithstanding any copyright notation herein.
\bibliographystyle{named}
\bibliography{ijcai20}
\end{document}